\documentstyle[12pt]{article}
\input{psfig}

\def\GeV{\,{\rm GeV}}
\def\TeV{\,{\rm TeV}}

\def\sec{\,{\rm sec}}
\def\Gyr{\,{\rm Gyr}}

\def\rcm{\,{\rm cm}}
\def\km{\,{\rm km}}

\def\Mpc{\,{\rm Mpc}}

\def\eV{{\,\rm eV}}

\def\cmm2{{\,\rm cm^{-2}}}
\def\cm2{{\,{\rm cm}^2}}
\def\cmm3{{\,{\rm cm}^{-3}}}
\def\gcmm3{{\,{\rm g\,cm^{-3}}}}
\def\kms{\,{\rm km\,s^{-1}}}

\def\mpl{{m_{\rm Pl}}}

\def\la{\mathrel{\mathpalette\fun <}}
\def\ga{\mathrel{\mathpalette\fun >}}
\def\fun#1#2{\lower3.6pt\vbox{\baselineskip0pt\lineskip.9pt
  \ialign{$\mathsurround=0pt#1\hfil##\hfil$\crcr#2\crcr\sim\crcr}}}

\begin{document}
\pagestyle{empty}
\begin{center}
\bigskip

\rightline{astro-ph/9704062}

\vspace{.2in}
{\Large \bf Ten Things Everyone Should Know About Inflation}
\bigskip

\vspace{.2in}
Michael S. Turner\\

\vspace{.2in}
{\it Departments of Astronomy \& Astrophysics and of Physics \\
Enrico Fermi Institute, The University of Chicago, Chicago, IL~~60637-1433}\\

\vspace{0.1in}
{\it NASA/Fermilab Astrophysics Center\\
Fermi National Accelerator Laboratory, Batavia, IL~~60510-0500}\\

\end{center}

\vspace{.3in}
\centerline{\bf ABSTRACT}
\bigskip
These lecture notes are organized into ten lessons
that summarize the status of inflationary cosmology.

\newpage
\pagestyle{plain}
\setcounter{page}{1}
\newpage

\section{Inflation is a Bold and Expansive Paradigm}
The hot big-bang cosmology is very successful.  It provides
a physical description of the Universe from about $10^{-2}\sec$
onward \cite{bbsuccess}.  However, it raises fundamental questions
about initial conditions:  The origin of the smoothness and flatness
of our Hubble volume, the small (one part in $10^5$) density inhomogeneities
needed to seed all the structure seen in the Universe today, and the
tiny baryon asymmetry that results in the existence of matter today.

Inflation explains how a region of size much, much
greater than our Hubble volume could have become smooth and flat \cite{guth}
as well as the origin of the density inhomogeneities needed
to seed structure \cite{scalar}.  With regard to the smoothness and flatness,
inflation is a temporary fix:  It does not guarantee that the observable
Universe in the exponentially
distant future will be isotropic and homogeneous \cite{nohair}.

Models of inflation are based upon well defined, albeit speculative
physics -- usually the semi-classical evolution of a weakly coupled
scalar field.  The physics is speculative because a) there is no
evidence for the existence of even a single fundamental scalar field and b)
the energy scale associated with inflation is typically much greater
than 1\,TeV and in most models around $10^{14}\GeV$.

I believe that it is fair to say that inflation has revolutionized
the way cosmologists view the Universe.  It leads to the current working
hypothesis for an extension of the
standard cosmology:  The Inflation/Cold Dark Matter Paradigm.  This paradigm
has the potential to extend the standard cosmology back to
times as early as $10^{-32}\sec$ and
address almost all the pressing questions in cosmology.  The
key elements of this paradigm are:  flat Universe, nonbaryonic dark
matter in the form of slowly moving elementary particles (cold dark
matter), and nearly scale-invariant, adiabatic density perturbations.
As I will emphasize, the inflation/cold dark matter paradigm is highly
testable and a flood of observations are doing so.  At the outside,
within the next decade this paradigm will have
been falsified or more firmly established.

There are even grander implications of inflation, albeit very
difficult to test \cite{eternal}.  Cosmologists have long used
the Copernican principle to argue that the entire Universe must be smooth
because of the smoothness of our Hubble volume.  In the post-inflation
view, our Hubble volume is smooth because it is a small
part of a region that underwent inflation, and thus it need not reflect
the large-scale features of the Universe as a whole.  On
the largest scales the structure of the Universe is likely to be
very rich:  Different regions may have undergone different amounts of
inflation, beginning at different times; some regions may not have undergo
inflation and may have collapsed to black holes; other regions
may be governed by different
realizations of the laws of physics because they
evolved into different vacuum states of equivalent energy.
It is likely that most of the volume of the Universe is still undergoing
inflation and that inflationary patches are being constantly produced
(eternal inflation).  In this case, ``the age of the Universe'' is
a meaningless concept:  Our expansion age merely measures the
time back to the end of our inflationary event.

If inflation is correct, it will be a major advance in our understanding
of the origin and evolution of our Hubble volume and it will open a new window
on physics beyond the standard model of particle physics.  It is
possible that inflation is just plain wrong, and over the years
other explanations have put forth to address the dilemma of the
initial data.  For example, Penrose has suggested the smoothness and flatness
of the Universe has to do with the nature of initial singularities
\cite{penrose}.  It has, however, been shown that
any microphysical solution to the horizon and flatness problems must
involve the two key elements of inflation -- superluminal expansion
and entropy production -- suggesting to me that inflation or something
closely related is likely to be the correct explanation \cite{htw}.

\section{There is No Standard Model of Inflation}

It would be nice if there were a standard model of inflation, but
there isn't.  Because inflation involves physics beyond the standard
model of particle physics and is probably to tied to fundamental
physics at energies of ${\cal O}(10^{14}\GeV )$ this is not surprising.
What is important, is that inflationary models make three robust
predictions (see next Section) which allow the
paradigm to be decisively tested.  Moreover, cosmological measurements
should also be able to discriminate between different
models (see final Section).

The two required elements of any inflationary model are:
superluminal expansion (i.e., accelerated expansion, $\ddot R >0$)
and massive entropy production \cite{htw}.
They are usually achieved by means of the classical evolution
of a scalar field rolling down its potential-energy curve.  During
the first part of its evolution, the field rolls so slowly that
its potential energy density is nearly constant; this drives a nearly
exponential expansion (superluminal expansion).
During the late part of its evolution,
the scalar field rapidly oscillates about the minimum of its potential
and the decay of these oscillations eventually leads to the production
of particles and the reheating of the Universe (entropy production).
The entropy produced is the heat that today is the Cosmic Background
Radiation (CBR).  Because of the massive entropy produced, any initial baryon
asymmetry is diluted to a level much, much less than the observed
${\cal O}(10^{-10})$ baryon number per photon and baryogenesis after
inflationary is mandatory.
The basic mechanics of inflation are well understood
and summarized in Ref.~\cite{kt}.

All models of inflation have one feature in common:  the scalar
field responsible for inflation has a very flat potential-energy
curve and is very weakly coupled.  This typically implies
a dimensionless coupling of the order of $10^{-14}$.
Such a small number, like other small numbers in physics (e.g.,
the ratio of the weak to Planck scales $\approx 10^{-17}$ or
the ratio of the mass of the electron to the $W/Z$ boson masses $\approx
10^{-5}$), runs counter to one's belief that a truly fundamental
theory should have no tiny parameters, and cries out for an
explanation.\footnote{It is sometimes stated that inflation
is unnatural because of the small coupling of the scalar field
responsible for inflation; while the small coupling certainly begs
explanation, inflationary models are not unnatural in
the technical sense as the small number is always stabilized
against the effect of quantum corrections.}
In some models, the small number in the
inflationary potential is related to other small numbers in
particle physics:  for example, the ratio of the electron mass
to the weak scale or the ratio of the unification scale to
the Planck scale.  Explaining the origin of the small number
associated with inflation is both a challenge and an opportunity.

Models of inflation range from the very simple
(e.g., chaotic inflation \cite{chaotic}) to those that attempt to be part of
a grander scheme (e.g., models that make
contact with speculations about physics at very high energies --
grand unification \cite{pi}, supersymmetry \cite{florida,olive,lbl},
preonic physics \cite{pati}, or supergravity \cite{liddle}.)
Some have attempted to link inflation with superstring theory \cite{banks};
others have focussed on the naturalness issue, trying to explain the small
dimensionless number associated with inflation \cite{natural}.

While the scale of the vacuum energy
that drives inflation is typically of the order of $(10^{14}\GeV)^4$, a
model of inflation at the electroweak scale, vacuum energy $\approx(1\TeV )^4$,
has been proposed \cite{knox}.  Multiple epochs of inflation are
also possible \cite{multiple}.  Inflation has been considered
in the context of alternative theories of gravity.
The most successful is first-order inflation \cite{lapjs,kolbreview},
where gravity is described by Jordan -- Brans -- Dicke theory (or a similar
theory of gravity) and inflation is triggered by a strongly first-order
phase transition (e.g., GUT symmetry breaking) of the kind originally
envisioned by Guth \cite{guth}.

There are certainly details of inflation that are both model-dependent and not
completely understood.   For example, the basics of reheating
were laid out early on \cite{basics}; however, important details are still
under study today \cite{reheat}.  While the physics issues such as
reheating and model building are important and interesting, they do
not affect the basic predictions of inflation that are crucial to its testing.
In the end, observations may give the best guidance about models and
even physics issues.

\section{Inflation Makes Three Robust Predictions}

Inflation theorists are very inventive and there are probably no
set of predictions that are common
to all models of inflation.  However, a theory
without definite predictions is not testable -- and is hardly a
theory at all (Mach's principle provides an interesting
case in point).  The philosopher of science Karl Popper argued
that the status of a scientific theory is tied to its vulnerability -- strong
theories constantly subject themselves to falsification.
I believe that inflation is a strong theory in the sense of Popper
and that it makes three predictions which allow it to be
falsified.  They are:

\begin{enumerate}

\item {\bf Flat universe.}  This is perhaps the most fundamental
prediction of inflation.  Through the Friedmann equation it
implies that the total energy density is always equal to the
critical energy density; it does not however predict the form (or forms)
that the critical density takes on today or at any earlier or later epoch.

\item {\bf Nearly scale-invariant spectrum of gaussian density perturbations.}
These density perturbations (scalar metric perturbations)
arise from quantum-mechanical fluctuations
in the field that drives inflation \cite{scalar};
they begin on very tiny scales
(of the order of $10^{-23}\,$cm) and are stretched to astrophysical size
by the tremendous growth of the scale factor during inflation (factor
of $e^{60}$ or greater).  Scale invariant refers to the fact that the
fluctuations in the gravitational potential
are independent of length scale; or equivalently that the
horizon-crossing amplitudes of the density perturbations are independent
of length scale.  While the shape of the spectrum of density
perturbations is common to all models, the overall amplitude is
model dependent.  Achieving density perturbations that are consistent
with the observed anisotropy of the CBR and large enough
to produce the structure seen in the Universe today requires a horizon
crossing amplitude of around $(\delta\rho /\rho )_H
\approx 2\times 10^{-5}$.  This is the most severe
constraint on inflationary models and leads to the small
dimensionless number associated with inflation.

\item {\bf Nearly scale-invariant spectrum of gravitational waves.}
These gravitational waves (tensor metric perturbations) arise during
inflation from quantum-mechanical fluctuations in the metric itself
and today have wavelengths from ${\cal O}(1\km )$ to the size of the
present Hubble radius and beyond \cite{tensor}.   Scale invariant here
refers to the fact that gravitational waves of all wavelength
cross the horizon with the same dimensionless strain amplitude.
Once again, the overall amplitude is model dependent (proportional
to the inflationary vacuum energy).  The uniformity
of the CBR provides a cosmological upper bound to the overall amplitude,
but unlike density perturbations, there is no cosmological
lower bound to the amplitude of gravity-wave perturbations.

\end{enumerate}

There are other interesting consequences of inflation that
are not generic.  For example, in models of first-order inflation,
where reheating occurs through the nucleation and collision of
vacuum bubbles, there is an additional, much larger amplitude, but
narrow-frequency-band spectrum of gravitational waves ($\Omega_{\rm GW}h^2
\sim 10^{-7}$) \cite{vacuumpop}.  Large-scale primeval magnetic
fields of interesting size can be seeded during inflation \cite{bfield}.
It is also possible to produce topological defects during
or near the end of inflation \cite{defect} or
isocurvature perturbations in a matter
component (axions \cite{isoax}, baryons \cite{isobar}, or something
else \cite{icdm}).  Such auxiliary predictions are interesting, but are
not part of the core predictions that can be used to falsify inflation.
On the other hand, they could prove very helpful in establishing inflation.

\section{Can Inflation Lead to an Open Universe?}

Yes, BUT!!

Whether or not flatness is a generic prediction of inflation has
been the topic of much debate recently.  I believe that flatness
should be taken as a firm prediction of inflation and I will
explain why.  If there is
one episode of inflation, solving the ``horizon''
problem and solving the ``flatness'' problem
(maintaining $\Omega$ very close to unity until the present epoch)
are linked geometrically by the simple expression \cite{tilt}
\begin{equation}
|\Omega_0 - 1| \la \left( {H_0^{-1}\over d_{\rm Patch}} \right)^2
\label{eq:linkage}
\end{equation}
where $d_{\rm Patch}$ is the present size of the inflationary
patch that our Hubble volume resides within, which is assumed to
have size $H^{-1}$ at the beginning of inflation.  If we make no assumption
about the smoothness of the Universe on superHubble scales before
inflation or about our location within our inflationary region,
solving the horizon problem requires $H_0^{-1} \ll d_{\rm Patch}$
and this implies $|\Omega_0 -| \ll 1$.

Open inflation requires that this linkage be evaded and
that the amount of inflation be tuned to a specific value.
The number of e-folds of inflation
$N$ is determined by the shape of the scalar-field potential,
\begin{equation}
N = {8\pi \over \mpl^2}\int_{\phi_i}^{\phi_f} {V(\phi ) d\phi \over
        V^\prime } \, .
\end{equation}
The value of $N$ required to achieve a given value of $\Omega$ today
depends upon the reheating temperature after inflation, the value
of $\Omega$ before inflation, and the temperature today,
\begin{equation}
N = {1\over 2}\ln \left[ {|\Omega_i-1| \over |\Omega_0 -1|} \right]
        + \ln \left[ {T_{\rm RH}T_0 \over \mpl H_0}\right] \, .
\end{equation}
The amount of inflation needed is linked to both the initial state
and the epoch of our existence and invites one to invoke the
anthropic principle.  I see this as a major step backward and
counter to the spirit of the inflationary program.

In any case, the simplest way to evade Eq.~(\ref{eq:linkage})
is to assume that the smooth patch that inflates has
an initial size that is ten or even hundred times larger than
the Hubble radius.  The more elegant way is to assume two epochs
of inflation, the first ending with the nucleation of a bubble
and second tied to the slow roll of a scalar field \cite{open}.
The open Universe resides within the bubble nucleated by the
first episode of inflation (which looks like an open
universe \cite{gott}) and is reheated by the second, slow-roll
epoch of inflation.

Open, double inflation in the context of eternal inflation
can trade tuning for a distribution of values of $\Omega_0$.
My hunch is that the distribution is likely to be
very strongly peaked, either around $\Omega_0=0$ or
$\Omega_0 =1$, rather than uniform.  The recent
work of Vilenkin and Winitzki suggests that this is the case \cite{vilenkin}.
                                                                 
The scientific question of the flatness of the Universe will
be answered, probably within the next five years by using the
fine-scale anisotropy of the CBR.  If the Universe is found
to be flat, I will score it as an important victory for
inflation.  If the flatness prediction is falsified
I will consider it a major defeat.  If the Universe is found not
to be flat, but other tests of inflation prove successful
(e.g., CBR anisotropy and/or gravitational waves),
I will be willing to take another look at open inflation.

\section{Inflation Implies Particle Dark Matter and Maybe More}

While inflation predicts a flat, critical-density Universe, it
sheds no light on the form that the critical density should take.
Cosmological observations have narrowed the possibilities.
Denote the fraction of critical density contributed
of all forms of energy density by $\Omega_0$; inflation predicts
$\Omega_0 = 1$.
Big-bang nucleosynthesis constrains the baryon density to
be well below the critical density:  $0.007h^{-2} \le \Omega_B  \le
0.024 h^{-2} < 0.10$ (for $h>0.5$) \cite{cst}, which
implies that most of the critical density must
be in a form other than baryons.
When the primeval deuterium abundance is pinned down by
a definitive determinations of D/H in high-redshift hydrogen clouds,
the baryon density will be pegged to a precision of around 10\% or so.
Tytler and his collaborators have made a very strong case
for a primeval deuterium abundance of D/H$\, \simeq 2.5 \times
10^{-5}$ \cite{tytler}, which implies that $\Omega_B h^2 \simeq 0.024$
or $\Omega_B \simeq 0.05 (0.7/h)^2$.

It is also known that:  most of the matter is dark
(luminous matter contributes less than 1\% of the critical
density, $\Omega_{\rm LUM} \simeq 0.003h^{-1}$) and the fraction
of critical density in matter that clusters exceeds 30\%, $\Omega_M
> 0.3$ \cite{omegamatter}.  Thus,
it follows that at least 25\% of the critical density should be
in the form of nonbaryonic matter in the form of particles,
$\Omega_{\rm nb particles} = \Omega_M -\Omega_B \ga 0.25$.
Particle physics has provided three very good candidates
whose relic abundance (if they exist) should be of the order of the
critical density:  an axion of mass around $10^{-5}\,$eV
\cite{axion}; a neutralino of mass between
10\,GeV and 500\,GeV \cite{jungman}; and a neutrino of
mass of the order of 10\,eV \cite{neutrino}.

All are well motivated:  the axion is a prediction of
the most promising solution to the strong-CP problem; the neutralino
is predicted by supersymmetric extensions of the standard model;
and essentially all attempts to unify the forces and particles of
Nature lead to the prediction that neutrinos have small,
but nonzero, masses.  In fact, these three particle dark matter
candidates are so well motivated that we should probably
take seriously the possibility that more than one
might contribute significantly to the matter density today.

Finally, it is possible that there is another, even more exotic
component, which is smoothly distributed and contributes up to 70\%
of the critical density, $\Omega_X = \Omega_0 -\Omega_M \la 0.7$.
The fact that evidence for $\Omega_M \sim 1$ is still lacking and
that a case is mounting for $\Omega_M\sim 0.3$ \cite{omegamatter},
suggests that inflationists should consider the possibility
of a smooth component seriously.  Candidates for such include vacuum
energy, tangled strings, and rolling scalar fields \cite{smooth}.

While Occam's Razor argues against a smorgasbord, Nature might enjoy
a more interesting meal, and inflation gives no guidance.

\section{Large-scale Structure from Quantum Fluctuations}

This is perhaps the most striking prediction of inflation, and
I believe, the motivation for Stephen Hawking's description of
the COBE DMR discovery of CBR anisotropy \cite{dmr}
as ``the most important discovery
of all time.''  I believe Hawking said this thinking
the COBE discovery might prove to be the first evidence
that the density perturbations that seeded all structure in
the Universe arose from quantum fluctuations during inflation.

Recall, scale invariant refers to density perturbations that
cross the horizon with the same amplitude, independent of length
scale.  Different scales cross the horizon at
different times, so the spectrum of density perturbations
today is not independent of scale.

Gaussian means that the density contrast,
$\delta \rho ({\bf x},\, t)/{\bar \rho}$, is
a gaussian random field, described fully by its two-point correlation
function, or equivalently by the power spectrum, which is the Fourier
transform of the correlation function and the square of the
Fourier transform of the density contrast.

Both scale invariant and gaussian are generic predictions
as they are linked to central features
of inflation.  Because the scalar field that drives inflation is very
weakly coupled, it behaves like a free field and its
fluctuations are gaussian \cite{scalar}.  The density perturbations are
proportional to the scalar-field fluctuations and hence they
too should be gaussian.  The deviation of the fluctuations from
scale invariance is related to the steepness of the scalar potential;
since the scalar field responsible for inflation
must take the 60 or so Hubble times
to evolve to the minimum of its potential in order to solve the
horizon/flatness problems its potential cannot be too steep.

The relationship between the inflationary potential and the power
spectrum of density perturbations today ($P(k) \equiv \langle |\delta_k|^2
\rangle$) is given by
\begin{eqnarray}
P(k) & = & {1024\pi^3\over75}\,{k \over H_0^4}\,
        {V_*^3\over\mpl^6{V_*^\prime}^2}
        \left( k\over k_* \right)^{n-1} T^2(k)  \nonumber \\
n -1 & = & -{1\over 8\pi}\left({\mpl V^\prime_* \over V_*} \right)^2
        + {\mpl \over 4\pi}\left( {\mpl V_*^{\prime}\over V_*} \right)^\prime
           \nonumber \\
{dn\over d\ln k} & = &
-{1\over 32\pi^2}\left({{m_{\rm Pl}}^3V_*^{\prime\prime\prime}\over V_*}\right)
        \left({{m_{\rm Pl}} V_*^\prime\over V_*}\right) \nonumber\\
     & & \hspace*{1cm} + {1\over 8\pi^2}
        \left({{m_{\rm Pl}}^2V_*^{\prime\prime}\over V_*}\right)
        \left({{m_{\rm Pl}} V_*^\prime\over V_*}\right)^2
        - {3\over 32\pi^2}\left({m_{\rm Pl}} {V_*^\prime\over V_*}\right)^4 \nonumber\\
T(q) & = & {\ln \left(1+2.34q \right)/2.34q \over
        \left[1+3.89q+(16.1q)^2+(5.46q)^3+(6.71q)^4\right]^{1/4}} \,,
\label{scalarps}
\end{eqnarray}
where $V(\phi )$ is the inflationary potential, prime denotes $d/d\phi$,
$V_*$ is the value of the scalar potential when the
scale $k_*$ crossed outside the horizon during inflation,
$T(k)$ is the transfer function which accounts for
the evolution of the mode $k$ from horizon crossing until the present,
$q = k/h\Gamma$, and $\Gamma\simeq \Omega_Mh$ is the
``shape'' parameter \cite{BBKS}.  It is very convenient
to chose $k_* = H_0$ so that $V_*$ is the value of the inflationary
potential when the scale that fixes the CBR quadrupole crossed
outside the horizon.  These expressions are
given to lowest-order in the deviation from scale invariance,
and assume a matter-dominated Universe today; the next-order
corrections are given in Ref.~\cite{lidtur} and the analogous
expressions which include the possibility of a cosmological
constant are given in Ref.~\cite{whtur}.

\begin{figure}[t]
\centerline{\psfig{figure=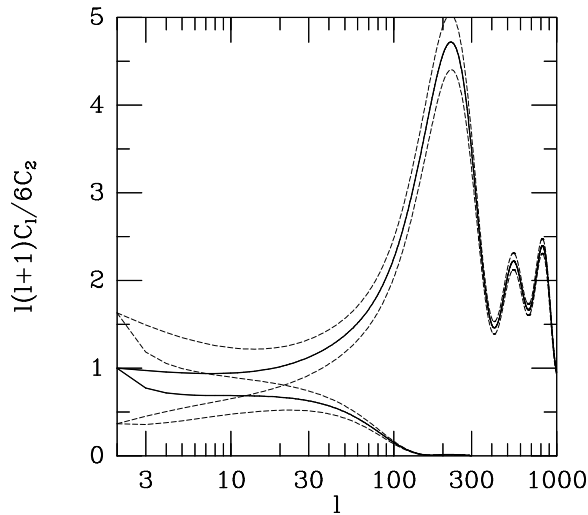,width=3in}}
\caption{Angular power spectra ($C_l \equiv
\langle |a_{lm}|^2\rangle$) of
CBR anisotropy for gravity waves (lower curves) and density perturbations
(upper curves), normalized to the quadrupole anisotropy;
broken lines indicate sampling variance.
Temperature fluctuations measured on angular scale
$\theta$ are approximately, $(\delta T/T)_\theta \sim
\protect\sqrt{l(l+1)C_l/2\pi}$
with $l\sim 200^\circ /\theta$ (courtesy of M.~White and U.~Seljak).}
\label{fig:staps}
\end{figure}

There are several important things to take note of:
\begin{enumerate}

\item The overall amplitude of the power spectrum depends upon the
combination $V_*^3/{V_*^\prime}^2$.

\item The quantity $n-1$, which measures the deviation from scale invariance,
is generally not equal to zero \cite{pjsmst}.

\item The deviation from scale invariance depends upon the steepness
of the potential ($\mpl V_*^\prime /V_*$) and the change in the steepness
\cite{mst,ll}.

\item Typically $n$ is less than 1, and for many models is in the
range $0.94$ to $0.96$ \cite{mst} (e.g., chaotic inflation \cite{chaotic}
and new inflation \cite{newinflation}).

\item There are inflationary models where $n$ is as larger than
1 (e.g., hybrid inflation \cite{hybrid}) and as small as 0.7
(e.g., power-law inflation and natural inflation).

\item Generally, the spectrum of perturbations is nearly, but not exactly,
a power law:  $dn/d\ln k$ is typically of the order of $-10^{-3}$;
only for power-law inflation is the spectrum an exact power law \cite{run}.

\item The shape of the transfer function, which determines the
level of inhomogeneity on small scales when the power spectrum
on large scales is normalized by COBE, depends upon $\Gamma = \Omega_M h$
(and to a lesser extent upon the baryon density \cite{baryontf}).

\end{enumerate}

Density perturbations give rise to CBR anisotropy which can be computed
very precisely \cite{wayne}.  CBR anisotropy probes the power spectrum
at early times ($z\simeq 1100$), when the perturbations were still in
the linear regime and astrophysical effects were minimal.  Thus, it
is one of the most important tests and powerful probes of inflation.
Expanding the CBR temperature on the sky in spherical harmonics,
\begin{equation}
{\delta T (\theta ,\phi ) \over T} =\sum_{lm} a_{lm}Y_{lm}(\theta ,\phi ),
\end{equation}
the anisotropy is fully characterized by its
angular power spectrum $C_l \equiv \langle |a_{lm}|^2\rangle$, shown in
Fig.~\ref{fig:staps}.  The ensemble average for the variance of the multipoles,
$\langle |a_{lm}|^2\rangle$, is related to the power spectrum (as described
in Ref.~\cite{wayne}); note, isotropy in the mean implies
$\langle a_{lm} \rangle = 0$.  The variance of multipole $l$ is dominated
by modes of wavenumbers $k\simeq lH_0/2$.
CBR anisotropy on large-angular scales ($l\ll 100$) arises
almost solely from the Sachs-Wolfe effect and to good approximation
can be computed analytically \cite{kt,sw}
\begin{equation}
C_l = {H_0^4\over 2\pi} \int_0^\infty (u/u_{\rm EQ})|j_l(u)|^2
     P(u/u_{\rm EQ}) du/u \, ,
\end{equation}
where $u=k\tau_0$ and $u_{\rm EQ} = k_{\rm EQ}\tau_0$.
The variance of the quadrupole anisotropy provides a handy means
of normalizing the power spectrum
\begin{equation}
S  \equiv {5\langle |a_{2m}|^2\rangle \over 4\pi} \simeq
         2.2\,{V_*/m_{\rm Pl}^4 \over (m_{\rm Pl} V_*^\prime /V_*)^2}.
\end{equation}

The detection of CBR anisotropy by the COBE DMR was a major
advance as it allowed the spectrum of density perturbations
to be normalized on very large scales ($k\sim H_0$).  For precisely
scale-invariant density perturbations and no gravitational-wave
contribution to CBR anisotropy the normalization procedure is
easy to describe:  $S$ is equated to the COBE determination of
the variance of the quadrupole anisotropy, $Q_{\rm COBE}
= (17\mu\,{\rm K}/2.2728\,{\rm K})^2
\simeq 3.8\times 10^{-11}$ \cite{cobe4yr}, which then implies
\begin{equation}
{V_*/m_{\rm Pl}^4 \over (m_{\rm Pl} V_*^\prime /V_*)^2}
        = 1.8 \times 10^{-11}
\end{equation}

Bunn et al. \cite{bunn} have done a careful analysis of the
COBE four-year data which takes account
of the fact that the COBE normalization for $S$ depends upon
$n$ as well as the possible contribution of gravitational waves
to CBR anisotropy.  (The ``pivot
point'' for the COBE data is $l\sim 15$; that is, the COBE
determinations of $C_{15}$ and $n$ are almost uncorrelated.)
This leads to the more accurate normalization
\begin{equation}
{V_*/m_{\rm Pl}^4 \over (m_{\rm Pl} V_*^\prime /V_*)^2}
        = 1.7\times 10^{-11} \,{\exp [-2.02(n-1)] \over
        \sqrt{1+{2\over 3}{T\over S}}}
\end{equation}
where $T$ is the tensor contribution to the variance of the CBR quadrupole.
The $1\sigma$ error is 15\%.
(Bunn et al. have also generalized this result to allow for the possibility of
a cosmological constant \cite{bunn}.)

The Bunn et al. normalization can also be expressed in terms of
the horizon-crossing amplitude for the comoving scale $k=H_0$:
\begin{equation}
\delta_H(k=H_0) \equiv \left[ {k^{3/2}|\delta_k| \over
        \sqrt{2\pi^2}} \right]_{k=H_0} = 1.9\times 10^{-5} \,
        {\exp [-1.01(n-1)]\over \sqrt{1+{2\over 3}\,{T\over S}}} .
\end{equation}

\section{Gravitational Waves:  The Smokin' Gun}

The inflation-produced gravitational waves are the smokin' gun
signature of inflation and crucial to learning about the
inflationary potential.
Both a flat Universe \cite{dickepeebles} and scale-invariant
density perturbations (so called Harrison--Zel'dovich spectrum
\cite{hz}) were discussed as features of any ``sensible cosmology'' long before
inflation.  The nearly scale-invariant spectrum of gravitational waves
which arises from quantum mechanical fluctuations excited in
the space-time metric during inflation is a very important prediction
of inflation that sets it apart from just any other sensible cosmology.
Detecting these gravitational waves will be very challenging.

Unlike the scalar perturbations, which must have an amplitude of
around $10^{-5}$ to seed structure formation, there is no astrophysical
clue as to the amplitude of the tensor perturbations.
They can be characterized by their power spectrum today \cite{gwperts}
\begin{eqnarray}
P_T(k)  & \equiv & \langle |h_k|^2\rangle
        =  {8\over 3\pi}{V_*\over \mpl^4}\left( k\over k_*\right)^{n_T-3}
        T_T^2(k) \nonumber \\
n_T & = & -{1\over 8\pi} \left( {\mpl V^\prime_* \over V_*} \right)^2
        \nonumber \\
{dn_T\over d\ln k} & = &
{1\over 32\pi^2}\left({{m_{\rm Pl}}^2V^{\prime\prime}\over V}
\right) \left( {{m_{\rm Pl}} V^\prime\over V}\right)^2  - {1\over 32\pi^2}
\left({{m_{\rm Pl}} V^\prime\over V}\right)^4
     =  -n_T[ (n-1) - n_T] \nonumber \\
T_T(k) & \simeq & \left[ 1+{4\over 3}{k\over k_{\rm EQ}} +{5\over 2}
\left( {k\over k_{\rm EQ}} \right)^2 \right]^{1/2} ,
\end{eqnarray}
where $T_T(k)$ is the transfer function for gravity waves
and describes the evolution of mode $k$ from horizon crossing
until the present, $k_{\rm EQ} = 6.22\times 10^{-2}\,{\rm Mpc^{-1}}\,
(\Omega_Mh^2/\sqrt{g_*/3.36})$ is the scale that crossed the horizon at
matter-radiation equality, $\Omega_M$ is the fraction of
critical density in matter, and $g_*$ counts the effective number
of relativistic degrees of freedom (3.36 for photons and three light
neutrino species).  The quantity $k^{3/2}|h_k|/\sqrt{2\pi^2}$
corresponds to the dimensionless strain (metric perturbation)
on length scale $\lambda =2\pi /k$.

Like density perturbations, gravity waves lead to
CBR anisotropy which can be fully described by an angular power spectrum.
The gravity-wave angular power spectrum
is related to the power spectrum and can be computed very
accurately \cite{cbrtensor}; it is shown in Fig.~\ref{fig:staps}.
The following analytical expression is accurate to about 10\%,
\begin{eqnarray}
C_l & = & 36\pi^2 {\Gamma (l+3)\over \Gamma (l-1)}\,\int_0^\infty\,
        F_l(u) (u/u_{\rm EQ})^3P_T(u/u_{\rm EQ})\,du/u \nonumber\\
F_l(u) & = & -\int^u_{(\tau_{\rm LS}/\tau_0)u} dy
        \left( {j_2(y)\over y}\right) \left({j_l (u-y)\over (u-y)^2 }\right) \\
\end{eqnarray}
where $u=k\tau_0$, $\tau_0=2H_0^{-1}$ is the conformal age of the
Universe today, and $\tau_{\rm LS}=\tau_0/\sqrt{(1+z_{\rm LS})}$ is the conformal
age at last scattering.
The tensor contribution to the variance of the CBR quadrupole is
a convenient normalization for the spectrum:
\begin{equation}
T  \equiv {5\langle |a_{2m}|^2\rangle \over 4\pi} =
        0.61 (V_*/m_{\rm Pl}^4) .
\end{equation}
The predicted variance of the CBR quadrupole anisotropy is $T+S$.

\begin{figure}[t]
\centerline{\psfig{figure=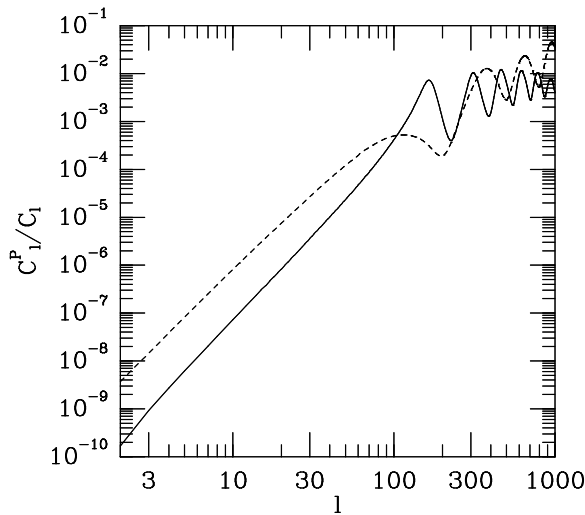,width=3in}}
\caption{Polarization angular power spectra for gravity waves
(broken) and density perturbations (solid).
The polarization of the CBR anisotropy is roughly
$\protect\sqrt{C_l^P/C_l}$ (courtesy of M.~White and U.~Seljak).}
\label{fig:polps}
\end{figure}

There are several important things to take note of:
\begin{enumerate}
\item The contribution of gravity waves to the variance of the
CBR quadrupole is proportional to the value of the vacuum energy
that drives inflation, and if $T$ can be determined, the energy
scale of inflation can be determined.

\item Using the COBE four-year results as an upper bound, $T<Q_{\rm COBE}$, it
follows that,  $V_* < 6\times 10^{-11} \mpl^4$, or equivalently,
$V_*^{1/4} < 3.4\times 10^{16}\GeV$.  This indicates that inflation,
if it occurred, involved energies much smaller than the Planck scale.
(To be more precise, inflation could have begun at a much higher
energy scale, but the portion of inflation relevant for us, i.e.,
the last 60 or so e-folds, occurred at an energy scale much smaller
than the Planck energy.  In chaotic inflation \cite{chaotic}, inflation is
supposed to begin at the Planck energy density.)

\item The four potential observables, $S$, $T$, $n-1$ and $n_T$,
depend upon three properties of the inflationary potential,
$V_*$, $V_*^\prime$ and $V_*^{\prime\prime}$.  Thus, the potential
and its first two derivatives can be expressed in terms of the
four observables with an additional consistency relation, $T/S = -7n_T$,
which is an important test of inflation \cite{consis}.  If $S$, $T$,
and $n-1$ can be determined, the potential and
its first two derivatives can be ``reconstructed''; in addition,
if $n_T$ can also be measured, the consistency of inflation can be tested
\cite{recons}.

\item The amplitude of the gravity-wave spectrum and its ``tilt'' (deviation
from scale invariance) are related:  the larger the amplitude, the
greater the tilt.  Moreover, this means the spectrum of gravity waves
can be described by a single parameter.

\item The tensor tilt, deviation of $n_T$ from 0, and the scalar
tilt, deviation of $n-1$ from zero, are in general not equal;
they differ by the rate of change of the steepness.
The only model where they are identical is power-law inflation.

\item The variation of the tensor index with scale, $dn_T/d\ln k$,
is typically ${\cal O}(10^{-3})$.

\end{enumerate}

There are two basic approaches to detecting the tensor perturbations:
CBR anisotropy and/or polarization and
direct detection of gravity waves.  The first approach probes the
spectrum at very long wavelengths, $\lambda \sim H_0^{-1}/100 - H_0^{-1}
\sim 10^{26}\rcm - 10^{28}\rcm$,
while the second probes much shorter wavelengths, $\lambda
\sim 10^{8}\,{\rm cm} - 10^{14}\,{\rm cm}$.  If some information
(detections/upper limits) could be
obtained at both wavelengths, both $T$ and $n_T$ could be measured or
at least constrained.

While the scalar and tensor angular power spectra
are very different (see Fig.~\ref{fig:staps}), sampling variance
sets a fundamental limit to how well the two can be separated from
measurements of the one CBR sky we have access to.
For multipole $l$, $2l+1$ multipole amplitudes can be used
to determine the variance; ``the variance of the variance'' is
\begin{equation}
{\langle (C_l^{\rm estimate} - C_l)^2\rangle^{1/2}\over
        C_l}  = \sqrt{2\over 2l+1} ,
\end{equation}
where $C_l^{\rm estimate}$ is the estimate based upon CBR measurements;
sampling variance is shown in Fig.~\ref{fig:staps}.
Using anisotropy alone, sampling variance implies that
the tensor contribution can be reliably separated
only if $T/S \ge 0.1$ \cite{knoxmst}.  The tensor and scalar perturbations
led to different levels of polarization of the anisotropy;
see Fig.~\ref{fig:polps}.
The sampling-variance limit based upon CBR polarization
is about a factor of five smaller \cite{knoxmst},
but requires that polarization on large-angular scales be measured
at less than a fraction of a percent.  Recently,
it has been pointed out that scalar and tensor perturbations
excite different patterns of polarization \cite{pattern}, which
could allow sampling variance to be evaded.  In any case,
the potential of polarization remains to be seen:
the signal is small (maximum polarization is a few percent of the anisotropy);
CBR polarization has yet to be detected; and the severity of
polarization foregrounds are yet to be determined.

Earth-based laser interferometers which operate in the 10\,Hz to kHz
range are being built in the US (LIGO) and in Europe (VIRGO).  A
space-based interferometer is being planned by ESA (LISA) and ideas
for a smaller mission are being discussed in the US (OMEGA).
Space-based interferometers could operate at frequencies as
low as $10^{-4}\,$Hz.

\begin{figure}[t]
\centerline{\psfig{figure=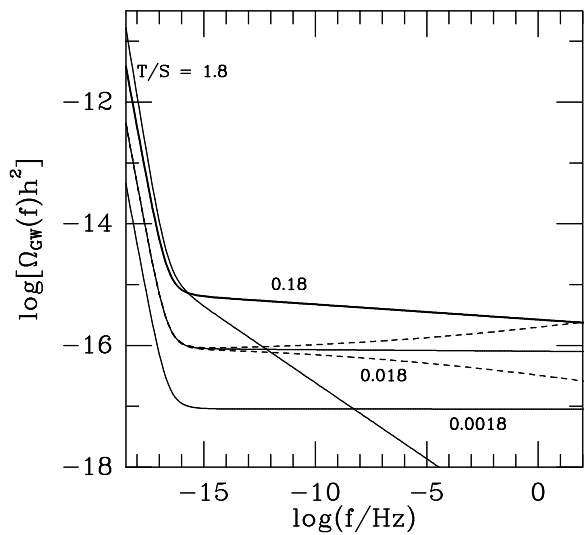,width=3in}}
\caption{Spectral energy density in gravity
waves produced by inflation; for $T/S = 0.018$,
$dn_T/d\ln k = -10^{-3}$, 0, $10^{-3}$.  $T/S =0.18$ (heavy curve)
maximizes the energy density at $f=10^{-4}\,$Hz.
Curves are from Eq.~(\protect\ref{eq:energy})
using $H_0=60{\rm km\,s^{-1}\,Mpc^{-1}}$, $\Omega_M =1$, and $g_*=3.36$
(from Ref.~\protect\cite{mstlisa}).}
\label{fig:gwspec}
\end{figure}

It is straightforward to go from the power spectrum to the
fraction of critical density contributed by gravity waves
per log frequency interval
\begin{equation}\label{eq:energy}
\Omega_{\rm GW}(f) \equiv {1\over \rho_{\rm crit}}\,
{d\rho_{\rm GW}\over d\ln k} = {\Omega_M^2\,V_*/ m_{\rm Pl}^4\,\over
(k/H_0)^{2-n_T}} \left[ 1+{4\over 3}{k\over k_{\rm EQ}} +{5\over 2}
\left( {k\over k_{\rm EQ}} \right)^2 \right] ,
\end{equation}
where $f = k/2\pi$ and $\Omega_{\rm GW}(f)$ is shown in
Fig.~\ref{fig:gwspec}.  The long plateau, frequencies greater
than $f_{\rm EQ}=k_{\rm EQ}/2\pi \sim 10^{-15}\,$Hz, reflects the
scale invariance of the gravitational-wave spectrum.  The rise for
smaller frequencies, as $f^2$, traces to the fact that
the longest wavelength modes crossed the horizon during
the matter-dominated epoch.  The energy density in gravitational waves
can also be expressed in terms of the {\it rms} metric perturbation
or strain, $h_{rms}^2(k)\equiv k^3|h_{\bf k}|^2/2\pi^2$,
\begin{equation}
\Omega_{\rm GW}(f) = {2\pi^2\over3} \,\left({f\over H_0}
\right)^2\,h_{rms}^2(k) = 6.3h^{-2} \times 10^{-7}\,(f/{\rm Hz})^2
\end{equation}

Using the relationship between the tensor spectral index $n_T$
and the amplitude $T$, and the COBE determination of the variance of the CBR
quadrupole anisotropy, Eq.~(\ref{eq:energy}) can be rewritten
in terms of $n_T$ (or $T/S$) alone \cite{mstlisa}.  Doing so,
it then follows that on the ``long plateau'' ($f\gg 10^{-15}\,$Hz)
\begin{eqnarray}\label{eq:sens}
\Omega_{\rm GW}(f)h^2 & = &
5.1\times 10^{-15} \,(g_*/3.36)\,{n_T\over n_T - 1/7} \nonumber \\
& \times & \exp [n_TN +{1\over 2} N^2 (dn_T/d\ln k) ],
\end{eqnarray}
where $N\equiv \ln (k/H_0)\simeq 33 +
\ln (f/10^{-4}{\rm Hz}) +\ln (0.6/h)$ and I have also allowed for
the possible variation of the tensor index $n_T$ which is
typically $-10^{-3}$.

The importance of the amplitude -- tilt
relationship can be seen in Fig.~\ref{fig:gwspec}.
Sadly, tilt goes in the direction of pushing $\Omega_{\rm GW}$
down as the amplitude $T/S$ is increased.  There is
a bright side:  $\Omega_{\rm GW}(f\sim {\rm Hz})$ is maximized
for $T/S \simeq 0.18$, which is close to the value predicted by
chaotic inflation and exceeds the sampling-variance
limit to the detection of tensor perturbations using CBR anisotropy alone.
While there are essentially no inflationary models
where $|n-1|\ll 0.1$, there are many models where $-n_T \ll 0.1$
(e.g., natural inflation and new inflation).  Because of the
amplitude -- tilt relationship, the gravity-wave background in
these models is very small.

\begin{figure}[t]
\centerline{\psfig{figure=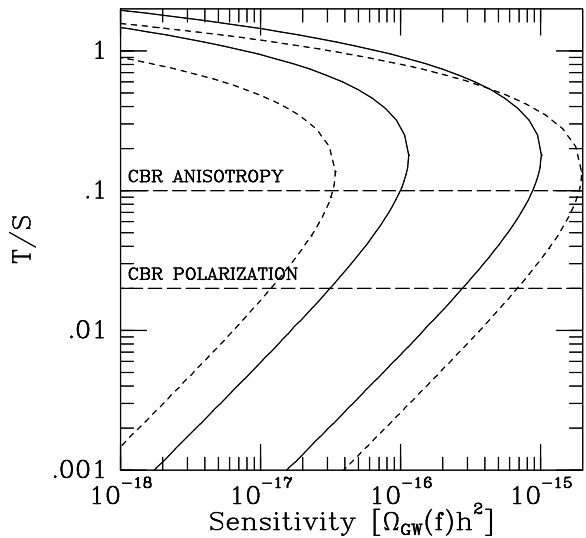,width=3in}}
\caption{The range of $T/S$ probed
(interval interior to parabola) as a function of energy
sensitivity for $f=10^{-4}\,$Hz
(solid curves) and $f=100\,$Hz (broken curves).  The
``pessimistic'' (left) parabola assumes $dn_T/d\ln k = -10^{-3}$
and the ``optimistic'' (right) parabola assumes $dn_T/d\ln k =
10^{-3}$.  Also shown are the sampling-variance limiting sensitivities for
CBR anisotropy and polarization (from Ref.~\protect\cite{mstlisa}).}
\label{fig:tssens}
\end{figure}

The range of $T/S$ accessible to a gravity-wave detector operating at
$f= 10^{-4}\,$Hz (appropriate for LISA) and $f= 100\,$Hz
(appropriate for LIGO and VIRGO) is shown as a function of the
detector energy sensitivity in Fig.~\ref{fig:tssens} \cite{mstlisa}.
A sensitivity of $\Omega_{\rm GW}(f)h^2 \sim 10^{-15}$ is needed for a
serious search for inflation-produced gravity waves.
With its initial strain detectors, the earth-based LIGO should be
able to detect a stochastic background of gravity waves with
90\% confidence provided $\Omega_{\rm GW}(f\sim 100\,{\rm Hz})h^2
\ge 2.8\times 10^{-6}$; with advanced strain detectors this
should improve to $2.8\times 10^{-11}$ \cite{ligo,ballen}.
Unfortunately, this misses the mark by four orders of magnitude.
(If one were to ignore the relation between tilt and amplitude and
assume $n_T\equiv 0$, LIGO would only miss by three orders
of magnitude \cite{ballen}.)

Because the energy density in gravity waves is proportional to
strain squared times frequency squared, a detector operating at
lower frequency has better energy-density sensitivity for fixed
strain sensitivity.  Earth-based detectors cannot
cannot operate at lower frequencies because of seismic noise, but
space-based detectors can.  The design study for the
space-based LISA indicates an energy sensitivity of
around $\Omega_{\rm GW}(f)h^2 \sim 10^{-13}$ at $f=10^{-4}\,$Hz \cite{lisa},
which is more promising, but still misses by two orders of magnitude.
(There is also a worrisome background of coalescing white-dwarf
binaries, which could dominate inflation at frequencies greater
than around $10^{-4}\,$Hz \cite{300yrs}.)

Detection of the inflation-produced gravity waves presents a very
great challenge.  But the possible payoffs are commensurately large:
A smokin' gun test of inflation; a determination of the energy scale
of inflation (through $T$); and a consistency test of inflation
(through $n_T$ and $T/S$).
                                               
\section{CDM:  A Testable, Ten Parameter Theory}

\begin{figure}[t]
\centerline{\psfig{figure=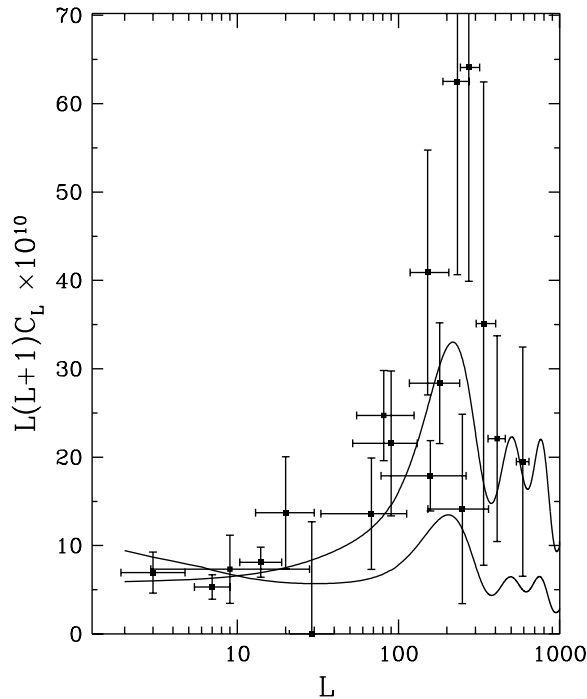,width=3in}}
\caption{Summary of CBR anisotropy measurements
and predictions for two CDM models.
Plotted are the squares of the measured multipole amplitudes ($C_l = \langle
|a_{lm}|^2\rangle$) versus multipole number $l$.
The relative temperature difference on angular scale $\theta$ is
given roughly by $\protect\sqrt{l(l+1)C_l/2\pi}$ with $l\sim 200^\circ /\theta$.
The theoretical curves are standard CDM (upper curve)
and CDM with $n=0.7$ and $h=0.5$ (lower curve) (from Ref.~\protect\cite{dgt}).}
\label{fig:cbrtoday}
\end{figure}

CBR anisotropy has been detected on angular scales ranging
from $100^\circ$ to a fraction of degree at the level of
about $\delta T/T \sim 10^{-5}$ (see Fig.~\ref{fig:cbrtoday}).
This provides strong support for the notion that
structure formed by the gravitational amplification of small
primeval density inhomogeneities.  One of the pressing
problems in cosmology is the formulation of a detailed and
coherent picture of structure formation.  The two key elements
in any such theory are the quantity and composition of matter in the Universe
and the nature of the perturbations that seed the formation of
structure.  Inflation makes definite predictions about
both, and with the rapidly growing number of high-quality
observations that bear on the issue, structure formation
has become an important testing ground for inflation.

Recall, inflation and astrophysical data indicate the following
about the quantity and composition of matter in the Universe:
baryons contribute a small fraction of the critical density,
$\Omega_B = (0.007 - 0.024)h^{-2}$; particle dark matter plus
baryons contribute at least 30\% of the critical density;
and there may be a smooth component that brings the total to
the critical density.  The inflationary prediction concerning
the seed perturbations is sharper:  almost scale-invariant gaussian
density perturbations, whose horizon-crossing amplitude is determined
by COBE to be about $2\times 10^{-5}$.

Particle dark matter can be classified by its velocity dispersion
at the epoch of matter -- radiation equality:  cold, $v_{rms} \ll
1$; hot, $v_{rms} \sim 1$; and warm, $v_{rms}$ not too much smaller
than 1.  Neutrinos and neutralinos were once in thermal equilibrium
and their velocity dispersion is set by the temperature at
matter -- radiation equality ($T_{\rm EQ} \simeq 6h^2\eV$)
and is inversely proportional to
their mass.  Neutrinos are light and therefore hot; neutralinos
are heavy and therefore cold.  Axions are cold in spite of their small
mass because they were never in thermal equilibrium and were produced
by a coherent, rather than thermal, process \cite{axion}.
Around the time of matter -- radiation
equality, density perturbations on small scales can be damped
by the freestreaming of dark matter particles from regions of
high density to regions of low density; for neutrinos this
is a significant effect and perturbations on scales less than
about 30\,Mpc are strongly damped.
For cold dark matter the freestreaming scale is much less than 1\,Mpc and
most likely uninteresting.  For warm dark matter the freestreaming scale
is around 1\,Mpc (essentially by definition) and has interesting
consequences.

For hot dark matter structure must form ``top down'' -- superclusters
form and fragment into galaxies.  More than a decade ago this possibility
was studied and was found to be wanting \cite{nohdm}.  Put
simply, there is every evidence that structure formed from
the bottom up:  The bulk of the galaxies formed at redshifts from
two to four and superclusters are only forming today.  Warm dark
matter is problematic because subgalactic-sized objects must form
from the fragmentation of galaxies; the abundance of high-redshift
(up to redshifts of almost five) hydrogen clouds is difficult at
best to reconcile with this fact \cite{nowdm}.
That leaves cold dark matter as the unique ``prediction'' of inflation.
As we shall see, this prediction has been very successful.

Here are the essential features of CDM \cite{blumetal,cdmsim}:
(1) it is hierarchical, with
smaller things forming first and larger things forming (slightly)
later; (2) because the amplitude of density perturbations on
very small scales varies slowly with scale,
$k^{2/3}|\delta_k| \propto \log k$ for $k\gg k_{\rm EQ}$, structure
formation is not strongly hierarchical; (3) in COBE normalized CDM
the first stars (in globular-cluster size objects) form at redshifts
of ten or so, galaxies begin forming at redshifts of five (with the
bulk forming between $z\sim 4$ and $z\sim 2$),
clusters begin forming at redshifts around one, and superclusters
are just becoming gravitationally bound today;
(4) CDM particles form the cosmic infrastructure on all scales --
in galaxies they are the dark halos and in clusters they are the
dark matter that pervades the cluster.

When the CDM scenario emerged more than
a decade ago many referred to it as a no-parameter
theory because it was so specific compared to previous
models for the formation of structure.  This was an overstatement
as there are cosmological quantities that must be
specified.  However, the data available did not require precise
knowledge of these quantities to begin testing the CDM paradigm.

Broadly speaking the parameters can be organized
into two groups \cite{dgt}.  First are the cosmological parameters:  the
Hubble constant $h$; the density of
ordinary matter, $\Omega_B h^2$; the amplitude
of the scalar perturbations, $S$, the scalar power-law index, $n$,
and the rate at which it varies, $dn/d\ln k$; the amplitude
of the tensor perturbations, $T/S$, and tensor power-law index, $n_T$.
The inflationary parameters fall into this category
because there is no standard model of inflation;
on the other hand, once determined they can be used to discriminate
among models of inflation.

The second group specifies the composition of invisible matter
in the Universe:  radiation, dark matter, and vacuum energy.
Radiation refers to relativistic
particles:  the photons in the CBR, three massless
neutrino species (assuming none of the neutrino species has
a mass), and possibly other undetected relativistic particles
(some particle-physics theories predict the existence of additional
massless particle species).  At present relativistic particles
contribute almost nothing to the energy density in the Universe,
$\Omega_R \simeq 4.2 \times 10^{-5}h^{-2}$; early on -- when
the Universe was smaller than about $10^{-5}$ of its present
size -- they dominated the energy content; the level of radiation
today is important as it determines when the transition from
radiation domination to matter domination took place and thereby
the shape of the transfer function (through $\Gamma$).

Dark matter could include other particle
relics besides CDM.  For example, each neutrino species has a number density of
$113\cmm3$, and a neutrino species of mass $5\eV$
would account for about 20\% of the critical density
($\Omega_\nu = m_\nu/90h^{2}\eV$).  Predictions for
neutrino masses range from $10^{-12}\eV$ to several MeV, and
there is some experimental evidence that at least one of
the neutrino species has a small mass \cite{numass}.
Finally, there is the cosmological constant.
Introduced and then abandoned by Einstein to prevent the
expansion of the Universe, and resurrected by Bondi, Gold and
Hoyle in 1948 to address an age crisis,
it is still with us.  In the modern context it corresponds
to an energy density associated with the quantum vacuum.  At present,
there is no reliable calculation of the value that the cosmological
constant should take \cite{cosmoconst}, and so its existence
must be regarded as a logical possibility, with its value to
be determined by observations.  (As mentioned earlier, there are
even more exotic possibilities for the smooth component \cite{smooth}.)

The original no-parameter CDM model, often referred to as standard
CDM \cite{blumetal}, is characterized by simple choices for the
cosmological and the invisible matter parameters:
precisely scale-invariant density perturbations ($n=1$),
$h=0.5$, $\Omega_B =0.05$, $\Omega_{\rm CDM}=0.95$;
no radiation beyond the photons and the three massless neutrinos; no
dark matter beyond CDM; and zero cosmological constant.
Standard CDM served its purpose well as the DOS 1.0 of
cosmology:  it focussed attention on a specific CDM model.

While inflation models predict that the shape of the spectrum is
approximately scale-invariant, the overall amplitude depends
on the particular inflationary model.  For standard CDM the overall amplitude
was fixed by comparing the predicted level of inhomogeneity with that
seen today in the distribution of bright galaxies.
Galaxy-number fluctuations in spheres of radius $8h^{-1}\Mpc$ are unity:
\begin{equation}
\sigma_r^2 \equiv \int_0^\infty {dk\over k}\ {k^3P(k)\over 2\pi^2}\,
 \left( {3j_1(kr)\over kr}\right)^2 = 1
\end{equation}
for $r=8\,h^{-1}$Mpc.  This normalization ($\sigma_8 = 1$)
corresponds to the assumption that light, in the form of bright galaxies,
traces mass.  Choosing $\sigma_8$ to be less than one means
that light is more clustered than mass and is a biased tracer of mass.
There is some evidence that bright galaxies are somewhat
more clumped than mass with biasing factor $b\equiv 1/\sigma_8
\simeq 1 - 2$; e.g., the number fluctuations of IRAS galaxies
on the $8h^{-1}\Mpc$ scale is less than one:, $\sigma_8({\rm IRAS})
= 0.69 \pm 0.04$ \cite{irassigma8}, implying that infrared selected
galaxies are less clustered than optically bright galaxies.

As discussed earlier, COBE changed the normalization procedure;
given the values of the CDM parameters and normalizing to
COBE $\sigma_8$ can be computed.  Further, an independent means
of determining $\sigma_8$, based upon the abundance of rich
clusters, has been developed; comparing this value to the that
computed from the COBE normalized spectrum now provides a check/constraint.
But I am getting ahead of myself.

Is a ten parameter theory testable?  With sufficient high-quality data
the answer is yes.  The standard model of
particle physics has at least nineteen parameters; not only has it
been tested, but most of the parameters have been determined, many to
better than 1\% precision.  In the next two Sections I hope to make the
case that inflation/cold dark matter can be tested with the same
decisiveness.  Much of my case will rely upon CBR anisotropy;
if 2000 or so multipoles can be measured to a precision
close to that dictated by sampling variance, I believe this ambitious
goal is achievable.

\section{Status of Inflation:  So Far, So Good}

The testing of inflation necessarily focuses on its
three robust predictions and their consequences.

\subsection{Flatness}

\begin{figure}[t]
\centerline{\psfig{figure=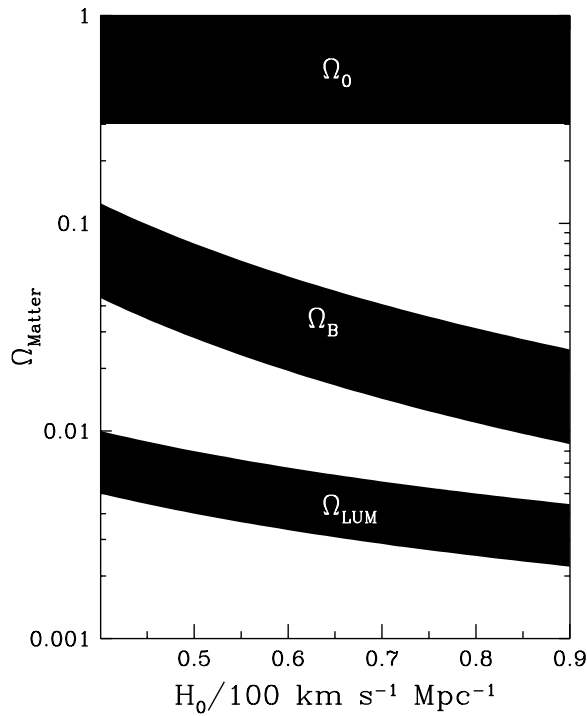,width=3in}}
\caption{Summary of knowledge of $\Omega_M$.
The lowest band is luminous matter, in the form of bright
stars and associated material; the middle band is the big-bang
nucleosynthesis determination of the density of baryons;
the upper region is the estimate based upon the peculiar
velocities of galaxies and other dynamical methods
\protect\cite{omegamatter}.  The gaps between the bands illustrate
the two dark matter problems: most of the ordinary matter is dark
and most of the matter is nonbaryonic (from Ref.~\protect\cite{dgt}).}
\label{fig:omega}
\end{figure}

The first prediction is a flat Universe with
nonbaryonic dark matter.  There is strong
evidence coming from a number of directions
that $\Omega_M$ is at least 0.3 \cite{omegamatter}.  This makes nonbaryonic
dark matter inescapable since the big-bang nucleosynthesis
upper bound is $\Omega_B < 0.024h^{-2} < 0.10$ (for $h>0.5$) and is
half way (on a logarithmic scale) to the simplest realization
of a flat Universe, $\Omega_M =1$; see Fig.~\ref{fig:omega}.
In the case that $\Omega_M \approx 0.3$
it is possible that the bulk of the closure density
resides in a smooth component.

Testing the flatness prediction has an even brighter future.
The position of the
first acoustic (Doppler) peak is sensitive to $\Omega_0$,
$l_{\rm peak} \simeq 220/\sqrt{\Omega_0}$.
The current data, while certainly not
definitive, put a smile on my face:  Hancock infers
$\Omega_0 = 0.7^{+1.0}_{-0.4}$ \cite{hancock}.
It is likely that even before MAP is launched in 2000, ground-based and
balloon-based measurements will determine the position of
the first acoustic peak well enough to peg $\Omega_0$ to 10\%.

Measurements of the deceleration of the Universe using the
magnitude -- redshift diagram for SNeIa constrain a nearly
orthogonal combination, $\Omega_M - \Omega_\Lambda$; together,
they can determine both $\Omega_M$ and $\Omega_\Lambda$.  (One should
keep in mind that there are more exotic possibilities for
the smooth component \cite{smooth,xcdm}.)
Assuming a flat Universe and using their first seven SNeIa, Perlmutter
et al \cite{perlmutter} derive the bound $\Omega_\Lambda <
0.51$ (95\%); or without the assumption of flatness,
$-0.4 < \Omega_M -\Omega_\Lambda < 2.7$ (95\%).
Perlmutter's group, the Supernova Cosmology
Project, now has a total of more than fifty SNeIa at redshifts
$z\sim 0.4 - 0.8$, and the High-Redshift Supernova Team has a similar
number; more definitive results should be coming soon.

\begin{figure}[t]
\centerline{\psfig{figure=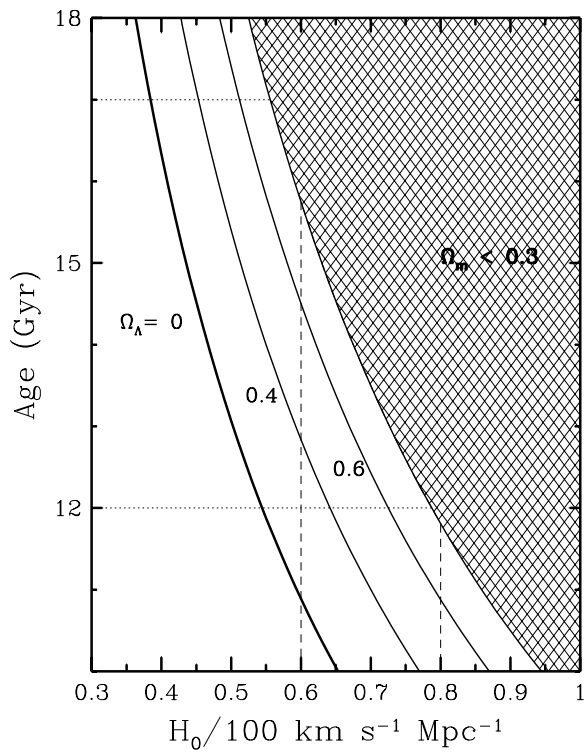,width=3in}}
\caption{The relationship between age and $H_0$ for flat-universe
models with $\Omega_{M} = 1 - \Omega_\Lambda$.
The cross-hatched region is ruled out because
$\Omega_{M} < 0.3$.  The dotted lines indicate
the favored range for $H_0$ and for the age of the Universe
(based upon the ages of the oldest stars) (from Ref.~\protect\cite{dgt}).}
\label{fig:agehubble}
\end{figure}

The combination of the age and Hubble constant can, in principle,
determine or at least constrain $\Omega_M$ and $\Omega_\Lambda$.
At the moment, the uncertainties preclude any firm
conclusions.  Taken at face value, the data seem to favor
a cosmological constant (if the Universe is flat); see
Fig.~\ref{fig:agehubble}.

A key consequence of the flatness prediction is the existence of
nonbaryonic dark matter.  The cold dark matter scenario won't be
fully tested until CDM particles are detected.  A large-scale
search for halo axions with sufficient sensitivity to detect them
(if they are there) is now underway \cite{axsearch},
and soon, CDMS, the Stanford --
Berkeley -- Case Western -- UCSB bolometric detector, will began
searching for halo neutralinos with sufficient sensitivity to detect
them for some models of low-energy supersymmetry \cite{cdms}.
SuperKamiokande and MACRO can search indirectly for neutralinos that
annihilate in the sun and produce high-energy neutrinos, and Ting's
AMS, which will be flown on the shuttle a year from now, will be
able to search for positrons and antiprotons produced by neutralino
annihilations in the halo.

\subsection{Gravity waves}
The search for inflation-produced gravitational waves was summarized
in Section 7.

\subsection{Density Perturbations/Cold Dark Matter}
Figure \ref{fig:cbrtoday} summarizes the
status of testing the second robust prediction of inflation
through CBR anisotropy:  The measurements are generally consistent
with the inflationary prediction.  The power-law
index is constrained to be $1.1\pm 0.2$, which is well within
the range predicted by inflation, and when COBE is used to
normalize the spectrum, there are CDM models that are consistent
with all the other observations.

\begin{figure}[t] 
\centerline{\psfig{figure=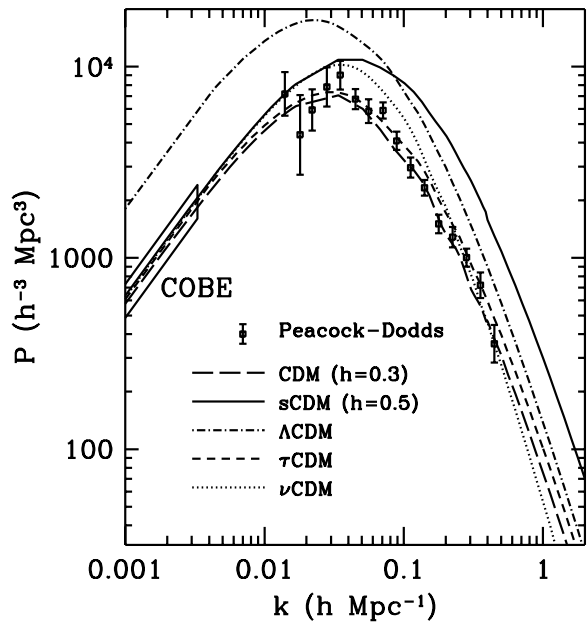,width=3in}}
\caption{Measurements of the power spectrum,
$P(k) = |\delta_k|^2$, and the predictions of different
COBE-normalized CDM models.  (COBE constrains the power
spectrum at wavenumbers $k$ around $2h\times 10^{-3}\Mpc$
as indicated by rectangle.)
The points are from several redshift surveys as compiled
by \protect\cite{pd}; the models are:  $\Lambda$CDM
with $\Omega_\Lambda =0.6$ and $h=0.65$; standard CDM (sCDM),
CDM with $h=0.35$; $\tau$CDM (with the energy equivalent
of 12 massless neutrino species) and $\nu$CDM with $\Omega_\nu = 0.2$
(unspecified parameters have their standard CDM values).
The offset between a model and the points indicates the level of biasing.
Note, $\Lambda$CDM does not pass through the COBE rectangle because
a cosmological constant alters the relation between the power spectrum
and CBR anisotropy (from Ref.~\protect\cite{dgt}).}
\label{fig:powerspec}
\end{figure}

There is much data that can be used to test the cold dark matter
scenario.  To focus the discussion, I will consider
four ``families'' of models, distinguished
by their invisible matter content: standard invisible matter content (CDM);
extra radiation ($\tau$CDM); small hot dark matter component ($\nu$CDM);
and cosmological constant ($\Lambda$CDM).
There are of course other possibilities:  extra radiation +
cosmological constant, or a more exotic smooth component,
which has been analyzed in Ref.~\cite{xcdm}.  Within each family, the five
cosmological parameters ($h$, $\Omega_Bh^2$, $n$, $T/S$ and
$n_T$) must be specified.  Once specified,
the power spectrum can be COBE-normalized and the expected level
of inhomogeneity in the Universe today computed.

In assessing the viability of CDM models I will summarize
work done in collaboration with Dodelson and Gates \cite{dgt};
others have done similar work \cite{othercdmstudies}.
We began with three robust observational constraints on the
power spectrum:  the shape of the power spectrum;
the power on cluster scales; and the early formation of objects.
The first constraint, the shape of the power spectrum on
scales from a few Mpc to a few 100 Mpc (see Fig.~\ref{fig:powerspec}),
comes from redshift surveys of the distribution of bright galaxies today
\cite{pd}.  In the absence of an understanding of the relationship
between the distribution of light, which is what these surveys
determine, and of mass, the bias factor is left
as a free parameter.  Models whose power spectra
deviates from the measured power spectrum (as compiled in
Ref.~\cite{pd}) by more than two sigma
(value of $\chi^2$ whose likelihood is less than 5\%)
were rejected.  (Very roughly, this constrains the shape
parameter:  $\Gamma =\Omega_Mh = 0.25 \pm 0.1$.)

The abundance of x-ray emitting clusters is sensitive to the
level of inhomogeneity on scales around $8h^{-1}\Mpc$ and
thus provides a good means of inferring the value of $\sigma_8$.  
Following \cite{sigma8cluster} we used $0.5 \le \sigma_8 \le 0.8 $ for
models with $\Omega_{M} = 1$ and
let this range scale with $\Omega_{}^{-0.56}$
for models with a cosmological constant ($\Omega_{\Lambda} = 1 - 
\Omega_{M}$) \cite{sigma8norm}.

The formation of objects at high redshift (early structure formation)
probes the power spectrum on small scales.  At
redshifts of two to four, hydrogen clouds, detected by their
absorption features in the spectra of high-redshift
quasars ($z\sim 4 - 5$), contribute a fraction of
the critical density, $\Omega_{\rm clouds} \simeq (0.001\pm 0.0002)h^{-1}$
\cite{Dlya}.  Insisting that the predicted level of inhomogeneity
is sufficient to account for this leads to a lower limit to the power
on small scales ($\lambda \sim 0.2h^{-1}\Mpc$).

\begin{figure}[t] 
\centerline{\psfig{figure=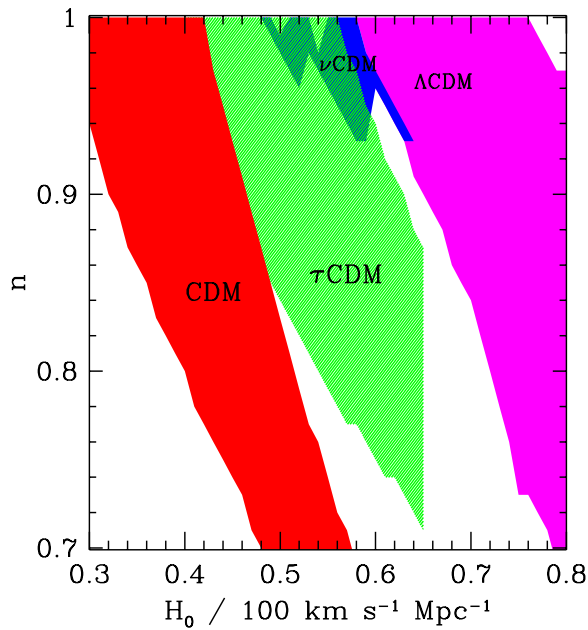,width=3in}}
\caption{Acceptable values of the cosmological
parameters $n$ and $h$ for CDM models with standard
invisible-matter content (CDM), with 20\% hot dark matter ($\nu$CDM),
with additional relativistic particles (the energy equivalent
of 12 massless neutrino species, denoted $\tau$CDM), and with a cosmological
constant that accounts for 60\% of the critical density ($\Lambda$CDM).
A model is considered viable if it passes
the three tests for {\it any} value of $\Omega_B h^2$
between $0.01$ and $0.02$ and any level of gravitational waves.
The $\tau$CDM models have been truncated at a Hubble constant
of $65\kms\Mpc^{-1}$ because a larger value would result in a
Universe that is younger than $10\Gyr$ (from Ref.~\protect\cite{dgt}).}
\label{fig:okcdmmodels}
\end{figure}

Figure \ref{fig:okcdmmodels} summarizes the viable models.
Models with standard invisible-matter content must lie in a
region that runs diagonally from smaller Hubble constant
and larger $n$ to larger Hubble constant and smaller $n$.
That is, higher values of the Hubble constant require more tilt.
As is well appreciated \cite{jpo,ll},
standard CDM is outside of the allowed range --
so much for DOS 1.0, onto Windows 95!
Current measurements of CBR anisotropy on the degree scale,
as well as the COBE four-year anisotropy data,
preclude $n$ less than about 0.7, which implies that the
largest $H_0$ consistent with the simplest CDM
models is slightly less than $60\kms\Mpc^{-1}$.

If the invisible-matter content is nonstandard, higher values of
$H_0$ can be accommodated.  With tilt and hot dark matter, $H_0$ as
large as $65\kms\Mpc^{-1}$ is consistent with the constraints.
The introduction of a cosmological constant permits
$H_0$ as large as $80\kms\Mpc^{-1}$, and additional radiation
allows a Hubble constant as large as the age constraint
permits (we assumed $t_0 \ge 10\,$Gyr, which requires $H_0\le
65 \kms\Mpc^{-1}$).

In passing I mention that a similar analysis
has been carried out for open-inflation models and they do
not fare nearly as well \cite{whitesilk}.  The
only viable models have $n>1.1$ or $\Omega_0 > 0.5$.  Cold dark
matter seems to be weighing in on the side of a flat Universe.

A host of other observations test CDM.  Some are more controversial
and/or open to interpretation.  They tend to distinguish the
cosmological-constant family of models from the other three families.
This is because models with standard invisible matter,
extra radiation, or a hot dark matter component are all
matter dominated today and have the same kinematic properties
-- age for a given Hubble constant and distance to an object of given
redshift.  The introduction of a cosmological constant leads to
an older Universe and greater distance to an object at fixed redshift.

Together, the Hubble constant and age of the Universe,
have great leverage.  Determinations of
the Hubble constant based upon a variety of techniques (Type Ia and II
supernovae, IR Tully-Fisher and fundamental-plane methods)
have converged on a value between $60\kms\Mpc^{-1}$ and $80\kms\Mpc^{-1}$
\cite{hubbleconst}.
This corresponds to an expansion age of less than $11\Gyr$ for a flat,
matter-dominated model; for $\Lambda$CDM, the expansion age can be
significantly higher, as large as $16\Gyr$ for $\Omega_\Lambda = 0.6$
(see Fig.~\ref{fig:agehubble}).  On the other hand, the
ages of the oldest globular clusters indicate that the Universe is
between $13\Gyr$ and $17\Gyr$ old; further, these age determinations,
together with the those for the oldest white dwarfs and the long-lived
radioactive elements, provide an
ironclad case for a Universe that is at least $10\Gyr$ old
\cite{age1,age2,age3}.  At face value, the age/Hubble constant
combination favor $\Lambda$CDM.  But again, I want to stress that, within
the uncertainties in both the age and Hubble constant, all of
the models are viable.

Clusters are large enough
that the baryon fraction should reflect its ``universal value,''
$\Omega_B/\Omega_{M} = (0.007 - 0.024)h^{-2}/(1-\Omega_\Lambda )$.
Most of the (observed) baryons in clusters are in the hot,
intracluster x-ray emitting gas.  From x-ray measurements of
the flux and temperature of the gas, baryon fractions
in the range $(0.04 - 0.10)h^{-3/2}$ have been
determined \cite{gasratio1,gasratio2,gasratio3}; further,
a recent detailed analysis and comparison to numerical models
of clusters in CDM indicates
an even smaller scatter, $(0.07\pm 0.007)h^{-3/2}$ \cite{evrard}.
From the cluster baryon fraction and $\Omega_B$,
$\Omega_{M}$ can be inferred:  $\Omega_{M}
= (0.25\pm 0.15)h^{-1/2}$, which for the lowest Hubble constant
consistent with current determinations ($h=0.6$) implies
$\Omega_{M} = 0.32 \pm 0.2$.  Unless one of the assumptions
underlying this analysis is wrong, $\Lambda$CDM is strongly favored.

E.~Turner emphasized the frequency of gravitational lensing of distant
QSOs as an important cosmological test \cite{elt}.
The underlying principle is simple:  the comoving distance to
fixed redshift depends upon the cosmology -- it is largest for
a cosmological constant, and in a matter-dominated Universe
decreases with increasing $\Omega_M$ -- and the probability of
lensing increases with comoving distance (more lenses along
the line of sight).
For a flat Universe, Kochanek has obtained the upper limit $\Omega_\Lambda
< 0.65$ (95\%), and for a matter-dominated Universe
$0.25 < \Omega_M < 2$ (95\%) \cite{kochanek}.  Neither result is decisive.

\begin{figure}[t] 
\centerline{\psfig{figure=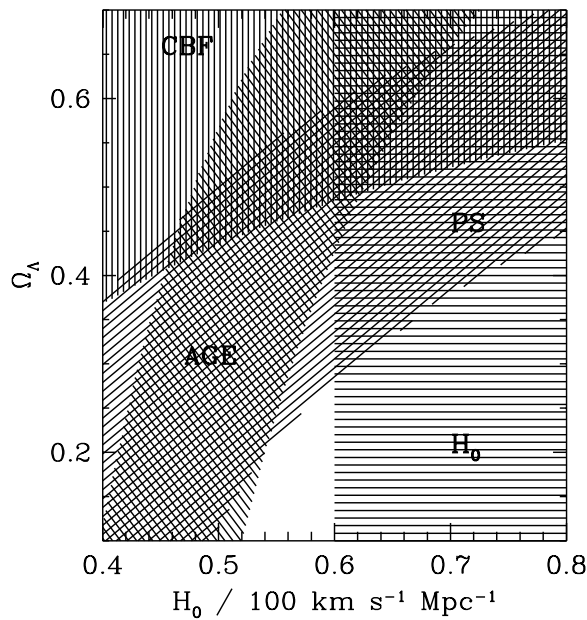,width=3in}}
\caption{Summary of constraints projected onto the $H_0$ --
$\Omega_{M}$ plane: (CBF) comes from combining the BBN
limit to the baryon density with x-ray observations of clusters; (PS)
arises from the power spectrum; (AGE) is based on age determinations
of the Universe; ($H_0$) indicates the range currently favored for the
Hubble constant.  The constraint $\Omega_\Lambda <0.66$ has been
implicitly taken into account since the $\Omega_\Lambda$ axis extends only
to 0.7. The darkest region indicates the parameters allowed by all
constraints.}
\label{fig:bestfit}
\end{figure}

$\Lambda$CDM is consistent with all the observations discussed
here as well as others; see Fig.~\ref{fig:bestfit}.  For this reason,
it is the strawman CDM model \cite{bestfit}.  The parameters
for this best fit model are:  $\Omega_\Lambda \sim 0.5 - 0.65$
and $h \sim 0.6 - 0.7$.  One should keep in mind that the introduction
of a cosmological constant is a big step, one which twice before
proved to be a misstep, and raises a fundamental question --
the origin of the implied vacuum energy, about $10^{-8}\eV^4$
\cite{cosmoconst}.

$\Lambda$CDM's hold on the title of best-fit CDM model is
by no means unshakeable:  should the Hubble constant turn out to
be less than $60\kms\Mpc^{-1}$ and should a flaw be found in the
cluster-baryon-fraction argument, the other models become very viable and
are theoretically more attractive.
Bartlett et al have pointed out that if the Hubble constant is
around $30\kms\Mpc^{-1}$, then CDM with $n\approx 1$ is consistent
with all the measurements discussed above \cite{hubble30}.
Lineweaver et al have analyzed the existing CBR anisotropy data
and conclude that it favors a Hubble constant of around this
value \cite{charley}.  The rub is squaring this
``determination'' of $H_0$ with the multitude of other determinations
that indicate a value almost twice as large.  Appealing to a difference
between the local value of the expansion rate and the global value
is of little help -- in the context of CDM, the difference, which
arises due to cosmic and sampling variance, is expected to be only
10\% or so \cite{hubblevariation}.

In finishing this status report I would like
to emphasize three things.  First, inflation is currently consistent
with all the observational data, which is no mean feat.  Second, cold dark
matter is consistent with a large body
of high-quality cosmological data, ranging from measurements of CBR
anisotropy to our growing understanding of the evolution of galaxies
and clusters of galaxies.
This too is no mean feat; at the moment, the only other potentially
viable paradigm for structure formation is topological defects + nonbaryonic
dark matter.  These models, when COBE normalized, appear to be in
great jeopardy as they predict very little power on small scales
(high bias).  Finally, the quantity of high-quality
data that bear on inflation and cold dark matter is growing rapidly,
leading me to believe that inflation/cold dark matter
will be decisively tested soon.

\section{The Future:  Precision Testing and More}

Inflation is a bold attempt to build upon the success of the standard
cosmology and extend our understanding of the Universe to times
as early as $10^{-32}\sec$ after the bang.  Its three robust predictions
-- flat Universe, nearly scale-invariant spectrum of density perturbations,
and nearly scale-invariant spectrum of gravitationally waves -- are
the keys to its testing.  In addition, much can be
learned about the specific, underlying model of inflation if other
measurements are made (e.g., the small anticipated deviation
from scale invariance and the level of gravitational waves).

\begin{figure}[t] 
\centerline{\psfig{figure=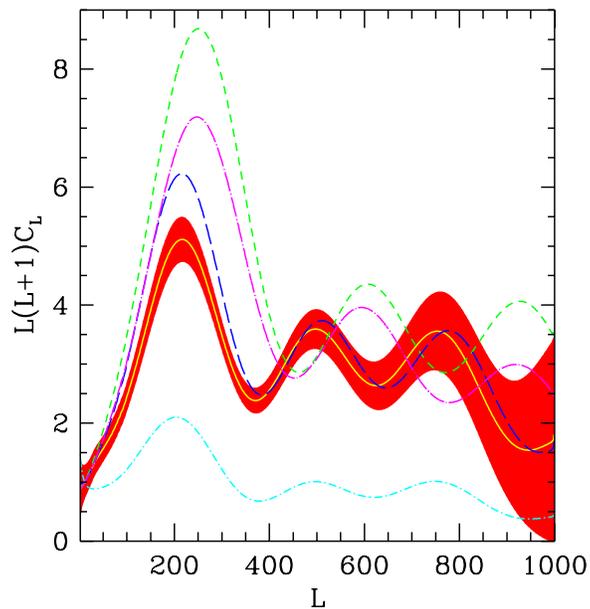,width=3in}}
\caption{Predicted angular power spectra of
CBR anisotropy for several viable CDM models and the anticipated
uncertainty (per multipole) from a CBR satellite experiment
similar to MAP.  From top to bottom the CDM models
are:  CDM with $h=0.35$, $\tau$CDM with the energy equivalent
of 12 massless neutrino species, $\Lambda$CDM with $h=0.65$ and
$\Omega_\Lambda = 0.6$, $\nu$CDM with $\Omega_\nu = 0.2$,
and CDM with $n=0.7$ (unspecified parameters have their
standard CDM values) (from Ref.~\protect\cite{dgt}).}
\label{fig:mapcbr}
\end{figure}

Cold dark matter, which is an important means of testing
inflation, is a ten-parameter theory,
$h$, $\Omega_Bh^2$, $S$, $n$, $dn/d\ln k$, $T/S$, $n_T$,
$\Omega_\nu$, $g_*$, and $\Omega_\Lambda$.  While this is a daunting
number of parameters, especially for a cosmological theory, there
is good reason to believe that within ten years the data will
overdetermine these parameters.  Crucial to achieving this goal
are the high-precision, high-resolution measurements
of CBR anisotropy that will be made over the next decade
by earth-based, balloon-borne and satellite-borne instruments
(see Fig.~\ref{fig:mapcbr}).  As a reminder of the power of
high-quality data, the standard model of particle physics has
nineteen parameters; precision measurements at Fermilab, SLAC,
CERN and other accelerator laboratories, as well as
nonaccelerator experiments, have both sharply tested the theory
as well as accurately determining the parameters.

Within five years we should be well on our way to precisely testing
inflation and cold dark matter.  In the next few years
ground-based and balloon-borne
anisotropy experiments (e.g., VSA, VCA, Boomerang, TOPHAT,
QMAX, and others) should be able to determine the approximate
position of the Doppler peak and thereby $\Omega_0$ to an accuracy
of around 10\%.  Because flatness is a fundamental prediction
of inflation, perhaps the most fundamental, this is a landmark
test.  On the same timescale, the Supernova Cosmology Project
and the High-Redshift Supernova Team will use the SNeIa magnitude --
redshift diagram to determine the deceleration of the Universe.
While the Doppler peak determines $\Omega_M + \Omega_\Lambda$,
the SNeIa measurement determines an almost orthogonal quantity,
$\Omega_M -\Omega_\Lambda$;
together, they can determine $\Omega_M$ and $\Omega_\Lambda$.

In the same time frame measurements of the Hubble constant
will play an important role.  Since the Universe is at least $10\Gyr$ old,
a definitive determination that the Hubble constant is $65\kms\Mpc^{-1}$
or greater would rule out all models but $\Lambda$CDM; on the other hand,
a determination that the Hubble constant is below $55\kms\Mpc^{-1}$
would undermine much of the motivation for $\Lambda$CDM.  The Hubble
Space Telescope calibration of secondary distance indicators such as SNeIa
with Cepheids in nearby galaxies and the maturation of physics-based
methods such as gravitational time delay and Zel'dovich -- Sunyaev are helping
to pin down $H_0$ more accurately \cite{newH0}.

There are other important tests that will be made on
a longer time scale.  The level of inhomogeneity in the
Universe today is determined mainly
from redshift surveys, the largest of which contains of
order $30,000$ galaxies.
Two larger surveys are underway.  The Sloan Digital Sky Survey
will cover 25\% of the sky and obtain positions for two hundred million
galaxies and redshifts for a million galaxies \cite{sdss}.
The Anglo -- Australian Two-degree Field is targeting
hundreds of two-degree patches on the sky and will obtain 250,000
redshifts \cite{2df}.  These two projects will determine the
power spectrum much more precisely and on scales
large enough ($500h^{-1}\Mpc$) to connect with measurements
from CBR anisotropy on angular scales of up to five degrees,
allowing bias to be probed.

The most fundamental element of cold dark matter --
the existence of the CDM particles -- will be tested
over the next decade.  Experiments with sufficient sensitivity to
detect the CDM particles that hold our own galaxy together if they are
in the form of axions of mass $10^{-6}\eV - 10^{-4}\eV$ \cite{axsearch} or
neutralinos of mass tens of GeV \cite{cdms} are now underway.
Evidence for the existence of the neutralino could also come
from particle accelerators searching
for other supersymmetric particles \cite{ellis}.
In addition, a variety of experiments sensitive to neutrino masses
are operating or are planned:
accelerator-based neutrino oscillation experiments at Fermilab,
CERN, and Los Alamos; solar-neutrino detectors in Japan, Canada,
Germany, Russia and Italy; and experiments at $e^+e^-$
colliders (LEP at CERN and CESR at Cornell)
to the study the tau neutrino.

The most telling test of inflation and cold dark matter will come
with the two new space missions that have recently
been approved -- MAP to be launched in 2000 by NASA and Planck
to launched by ESA in 2005.  Each will map CBR anisotropy over
the entire sky with more than thirty times the angular resolution of COBE
(resolution of $0.2^\circ$ for MAP and $0.1^\circ$ for Planck).
MAP should determine the angular power spectrum
out to multipole number 1000, and Planck out
to multipole number 2000, each to a precision close to that
dictated by sampling variance alone.
Theoretical studies \cite{learn} indicate that the results of
Planck should be able to determine $n$ to a precision
of less than one percent, $\Omega_\Lambda$ to a few percent,
$\Omega_0$ to less than one percent, $\Omega_\nu$ to
enough precision to test $\nu$CDM \cite{US}, the baryon density to
less than ten percent, and even the Hubble constant to one percent.

Inflation and cold dark matter are a bold attempt to extend
our knowledge of the Universe to within $10^{-32}\sec$
of the bang.  The number of observations that are testing the
cold dark matter theory is growing fast, and prospects for not
only testing the theory, but also discriminating among the
different CDM models and models of inflation are excellent.
If cold dark matter is shown to be correct, an important aspect of
the standard cosmology -- the origin and evolution of structure --
will have been resolved and a window to the early moments of the Universe
and physics at very high energies will have been opened.

While the window has not been opened yet, I would like to
end with one example of what one could
hope to learn.  As discussed earlier, $S$, $n-1$, $T/S$ and
$n_T$ are related to the inflationary potential and its first
two derivatives.  If one can measure the power-law
index of the scalar perturbations and the amplitudes of the scalar
and tensor perturbations, one can recover the value of the potential
and its first two derivatives around the point on the potential
where inflation took place \cite{recons}:
\begin{eqnarray}
V_* & = & 1.65 T\, {m_{\rm Pl}}^4  , \\
V_*^\prime & = & \pm \sqrt{{8\pi \over 7}\,{T\over S}}\, V_*/{m_{\rm Pl}} , \\
V_*^{\prime\prime} & = & 4\pi \left[ (n-1) + {3\over 7} {T\over S} \right]\,
V_* /{m_{\rm Pl}}^2 ,
\end{eqnarray}
where the sign of $V^\prime$ is indeterminate (under $\phi \leftrightarrow
-\phi$ the sign changes).  If the tensor spectral index
can also be measured the relation, $T/S = -7 n_T$,
can be used to test the consistency of inflation.
Reconstruction of the inflationary scalar potential would
shed light both on inflation as well as physics at energies of the
order of $10^{14}\GeV$.

\paragraph{Acknowledgments.}
Much of the material in these lectures derives from collaborative
work with Scott Dodelson, Evalyn Gates, Lloyd Knox, Andrew Liddle,
and Martin White.  I thank my collaborators for teaching me so much.
This work was supported by the DoE (at Chicago
and Fermilab) and by the NASA (at Fermilab by grant NAG 5-2788).

\end{document}